\documentclass[twocolumn,showpacs,showkeys,preprintnumbers,prd,superscriptaddress,nofootinbib]{revtex4-1}

\usepackage{amsmath}
\usepackage{amsfonts}
\usepackage{amssymb}
\usepackage{graphicx}
\usepackage{color}
\usepackage[]{hyperref}
\usepackage{booktabs}
\usepackage{cleveref}
\usepackage{subfigure}

\Crefname{equation}{Eq.}{Eqs.}
\Crefname{figure}{Fig.}{Figs.}
\Crefname{section}{Sec.}{Secs.}

\begin{document}

\title{Thermodynamic parametrization of dark energy}

\author{Salvatore Capozziello}
\email{capozziello@na.infn.it}
\affiliation{Dipartimento di Fisica ``E. Pancini", Universit\`a di Napoli ``Federico II", Via Cinthia 9, 80126 Napoli, Italy.}
\affiliation{Scuola Superiore Meridionale, Largo S. Marcellino 10, 80138 Napoli, Italy.}
\affiliation{Istituto Nazionale di Fisica Nucleare (INFN), Sez. di Napoli, Via Cinthia 9, 80126 Napoli, Italy.}
\affiliation{Laboratory  for Theoretical Cosmology, International Centre of Gravity and Cosmos,
Tomsk State University of Control Systems and Radioelectronics (TUSUR),
634050 Tomsk, Russia.}

\author{Rocco D'Agostino}
\email{rocco.dagostino@unina.it}
\affiliation{Scuola Superiore Meridionale, Largo S. Marcellino 10, 80138 Napoli, Italy.}
\affiliation{Istituto Nazionale di Fisica Nucleare (INFN), Sez. di Napoli, Via Cinthia 9, 80126 Napoli, Italy.}

\author{Orlando Luongo}
\email{orlando.luongo@unicam.it}
\affiliation{Dipartimento di Matematica, Universit\`a di Pisa, Largo B. Pontecorvo, 56127 Pisa, Italy.}
\affiliation{Universit\`a di Camerino, Divisione di Fisica, Via Madonna delle carceri, 62032 Camerino, Italy.}
\affiliation{NNLOT, Al-Farabi Kazakh National University, Al-Farabi av. 71, 050040 Almaty, Kazakhstan.}

\begin{abstract}

We  propose a  new parametrization of dark energy motivated by thermodynamics. To this aim, we consider Pad\'e polynomials to reconstruct the form of  deceleration parameter adequate to describe different epochs of  cosmic history and  divergence-free in the far future. The proposed scenario also fulfills  the demand of structure formation and contains the $\Lambda$CDM model as a limiting case. Thus, a numerical analysis at both background and perturbation levels is performed through the Markov Chain Monte Carlo method in view of the most recent cosmic data. We then use the observational constraints to explore the features of  dark energy evolution and compare our findings with the predictions of the standard cosmological model.

\end{abstract}


\maketitle

\section{Introduction}

The successful results of the standard cosmological model are based on  highly-precise measurements, as {\it e.g.}  cosmological abundances, indicating  the relative amounts of  the energy-matter  budget  \cite{uno1,uno2,uno3}. Even though the model turns out to be  predictive in describing early and late times, it is now being somewhat jeopardised by the discovery of several cosmological tensions \cite{due1}. Nevertheless, disclosing  the dark energy and dark matter natures remains a crucial challenge for modern cosmology \cite{due2}, implying as a consequence that  the standard paradigm, namely the $\Lambda$CDM model, may be somehow incomplete to describe the overall cosmic dynamics. For a different perspective, see \emph{e.g.} \cite{due3,due4}. Thus, characterising the Universe evolution without postulating a cosmological model \emph{a priori} \cite{cosmography1,cosmography2} might shed light on the dark sector properties whose dynamics could be addressed also by improving the geometry (see \textit{e.g.} \cite{Bamba, Oikonomou, Martino, Piedipalumbo}). In addition, requiring thermodynamics to hold throughout the Universe evolution is dutiful for any cosmological model that aims at extending the $\Lambda$CDM scenario \cite{Pavon12}. The validity of thermodynamics has been widely-investigated in cosmological scenarios, being a promising tool to reveal the dark sector nature \cite{tre1}. In this respect, thermodynamics could single out the most suitable benchmark to describe the overall Universe dynamics \cite{tre2,tre2bis,tre3}. 

In this work, we reconstruct both late and early time evolution of the Universe by virtue of  thermodynamic requirements over the cosmological kinematics. To do so, we assume the second law of thermodynamics to hold and the corresponding consequences on the apparent horizon in a spatially-flat Friedman-Lema\^itre-Robertson-Walker (FLRW) metric.  Moreover, we constrain the deceleration parameter evolution by assuming two physical conditions on the entropy $S$: first, it  cannot decrease, \emph{i.e.} $dS \geq 0$, with respect to an   evolutionary parameter.  Secondly, it must be a convex function  towards the last stages of the evolution, {\it i.e.}   $d^2S<0$ at  $z\rightarrow-1$. Consequently, it is possible to fix limits on how the Universe evolves at the present, early and future epochs. To this end, we build up a widely-general and suitable parametrization of the deceleration parameter, constructed by means of rational Pad\'e approximants and guaranteeing  stability at high redshifts. The $(0,1)$ Pad\'e series is selected to feature the deceleration parameter without postulating any cosmological model \emph{a priori}. Bearing this in mind, we infer the dark energy properties, highlighting the main differences with respect to the standard cosmological scenario. In particular, consequences on the background evolution are discussed in view of a numerical analysis performed on the most recent, model-independent observations at low redshifts. Furthermore, the analysis of cosmological perturbations is carried out through the growth rate of matter density fluctuations at high redshifts. 

The outline of the paper is as follows. In \Cref{sec:thermodynamics}, we introduce the constraints placed by the laws of thermodynamics on the cosmic evolution and the observational evidences of structure formation in the Universe.
In \Cref{sec:reconstruction},  a  dark energy parametrization is proposed by means of rational series, in agreement with thermodynamic and cosmological requirements. We thus test the proposed scenario against cosmic data in \Cref{sec:observations}, and we discuss the physical properties in lieu of the consequences on the dark energy evolution. Finally, \Cref{sec:conclusion} is devoted to conclusions and perspectives.

\section{Thermodynamics in the homogeneous and isotropic Universe} 
\label{sec:thermodynamics}

The use of thermodynamics in modern cosmology has reached great consensus since its validity is universal \cite{altro1}. Einstein's field equations fulfill standard thermodynamic  requirements and thermodynamic models of dark energy have been recently widely discussed and proposed as valid alternatives to the standard cosmological puzzle \cite{altro2}. On the other side, it has been argued that  deceleration parameter and its variation can be worked out independently of the underlying cosmological model \cite{altro3}. Thus, combining those two recipes, namely thermodynamics with kinematic reconstructions, can shed light on how dark energy might evolve at present, future and past epochs. 

Below, we present  thermodynamic and kinematic reconstructions considering cosmological entropy and  deceleration parameter. The two approaches are matched and combined with effective dark energy properties to represent the Universe expansion history without postulating  \emph{a priori} any cosmological model.

\subsection{Constraints from cosmological entropy}

The second law of thermodynamics establishes the tendency of physical systems to spontaneously approach an equilibrium state characterized by maximum entropy. 
This means that the entropy of macroscopic, isolated systems is increasing with time, namely $dS\geq 0$. Moreover, entropy is a convex function as evolving towards its last evolution phase. In the cosmological context, the latter property must hold for the redshift variable tending to the far future, \emph{i.e.} $d^2S<0$ for $z\rightarrow -1\,$.

In FLRW spacetime, late-time entropy is dominated by the entropy of the causal horizon, within which we consider the physically relevant part of the Universe to be. As shown in \cite{Wang06}, the appropriate boundary surface, where the laws of thermodynamics are satisfied, is the apparent horizon. This is given as 
	\begin{equation}
	\tilde{r}_h=(H^2+ka^{-2})^{-1/2}\,,
	\end{equation}
where $H\equiv \dot{a}/a$ is the Hubble parameter, $a=(1+z)^{-1}$ the cosmic scale factor, given in terms of redshift $z$,  and $k$ the spatial curvature. Neglecting quantum corrections, the entropy of the apparent horizon is proportional to its area, namely 
	\begin{equation}
	S_h \propto \mathcal{A}_h=4\pi \tilde{r}_h^2\,.
	\end{equation}
Therefore, the constraints from the second law of thermodynamics read $\mathcal{A}_h'\geq 0$ for any time, and $\mathcal{A}_h''<0$ at late times, where the prime indicates derivative with respect to the scale factor. These conditions can be used to constrain the cosmological dynamics.

For this purpose, we consider a spatially-flat Universe, $(k=0)$, and the deceleration parameter, defined as 
	\begin{equation}\label{qu}
	q\doteq-\dfrac{\ddot{a}a}{\dot{a}^2} =-1-\dfrac{\dot{H}}{H^2}\,,
	\end{equation}
where the dot denotes derivative with respect to the cosmic time.
We thus find
	\begin{equation}
	\dfrac{\mathcal A_h'}{\mathcal A_h}=\dfrac{2}{a}(1+q)\,, 
	\end{equation}
which satisfies the condition $\mathcal{A}_h'\geq 0$ if $q\geq-1$\,. On the other hand, we have
	\begin{equation}
	\dfrac{\mathcal A_h''}{\mathcal A_h}=\dfrac{2}{a}\left[q'+\dfrac{2q(1+q)}{a}\right], 
	\end{equation}
whose asymptotic behaviour for $a\rightarrow \infty$ is $\mathcal A_h''\sim 2\mathcal A_h \frac{q'}{a}$. Hence, in this limit we must have $q'<0$, and $q\rightarrow -1$ with $\frac{dq}{dz}>0$ as $z\rightarrow-1$.

\subsection{Constraints on the deceleration parameter}

Besides thermodynamic constraints, any possible scenario modelling cosmological dynamics should account for the observational predictions of cosmic structure formation. This implies that the behaviour of  deceleration parameter has to  obey a few asymptotic conditions and so, in the literature, it is often assumed that $q\rightarrow \frac{1}{2}$ for $z\gg 1$, to guarantee structure formation at high redshifts. 
However, this turns out to be not always accurate as, if no scalar fields enter the energy-momentum tensor, the early-time radiation contribution cannot be neglected so easily. 

To better clarify this fact, let us consider a generic dark energy model, in which a barotropic fluid is responsible for the evolution of dark energy \cite{algo1,algo2}:
	\begin{equation}
    E(z)^2= \Omega_{r0}(1+z)^4+\Omega_{m0}(1+z)^3+\Omega_{de,0}X(z)\,,
	\end{equation}
where $E(z)\equiv H(z)/H_0$ is the reduced Hubble parameter. Here, $X(0)=1$ and $\Omega_{de,0}=1 - \Omega_{m0} - \Omega_{r0}$, with $\Omega_{r0}$ and $\Omega_{m0}$ being the present-day values of radiation and matter density parameters, respectively.
Under the standard hypothesis in which dark energy is mainly a function of the inverse powers of $a(t)$, namely $X(a)\propto a^{-p}$, with $p\leq 4$,  we can see that radiation 
is the dominating species for $z\rightarrow\infty$. 

From \Cref{qu}, one then finds
	\begin{equation}\label{qugenerico}
	q=-1+\frac{(1+z)^3 [3 \Omega_{m0} + 4 \Omega_{r0} (1+z)]  \Omega_{de,0}(1+z) \frac{dX}{dz}}{
	2 (1 + z)^3 (\Omega_{m0} + \Omega_{r0} + \Omega_{r0} z) + 2 \Omega_{de,0} X(z)}\,.
	\end{equation}
Easy calculations show that $q\rightarrow1$, in the limit $z\rightarrow\infty$. Therefore, we can conclude that the asymptotic condition $q\rightarrow \frac{1}{2}$ is valid only up to the epoch of structure formation, after which the contribution of radiation cannot be neglected any further.

Bearing in mind these considerations, we  summarize the conditions over the deceleration parameter as follows:
	\begin{subequations}
	 \begin{align}
	& q\rightarrow -1 \ \left(\dfrac{dq}{dz}>0\right), \quad \text{as}\, \, z\rightarrow -1 \label{constraint1}\\
	& q\geq-1\,, \quad  \forall z \label{constraint2} \\
	& q\rightarrow \dfrac{1}{2}\,, \quad  \text{for}\, \,  z\gg 1 \label{constraint3}
	\end{align}
	\end{subequations}

\section{Reconstruction of the cosmic history}
\label{sec:reconstruction}

Let us now consider a Universe filled with pressureless matter and dark energy only, governed by the Friedman equations\footnote{In the following, we use units such that $8\pi G=1$.}
	\begin{align}
	&H^2=\dfrac{1}{3}\left(\rho_m+\rho_{de}\right), \\
	&3H^2+\dot{H}=- p_{de}\,.
	\end{align}
Focusing on the dark energy evolution, we are here interested in constructing a dark energy parametrization that satisfies the aforementioned thermodynamic and kinematic constraints. 

To do so, we  make use of rational series under the form of well consolidated Pad\'e approximants \cite{Pade1982} over the deceleration parameter, \emph{i.e.} on the only quantity that can be written in agreement with conditions \eqref{constraint1}-\eqref{constraint3}.

The technique of Pad\'e approximants have been proved a very useful and trustworthy tool in cosmographic analyses \cite{cosmography2}, in which  the expansion history of the Universe can be investigated in a model-independent way through a set of kinematic variables \cite{Visser}. 

In particular, we work out the Pad\'e approximant of the deceleration parameter by virtue of the following considerations:

\begin{itemize}
    \item[-] the Pad\'e series enables one to go through the cosmic history regardless possible transition between different phases;
    \item[-] the Pad\'e series extends the Taylor polynomials and is stable at very large redshifts, avoiding divergences as $z\rightarrow \infty$, since the series itself is rational. See \cite{cosmography2} for details.
    \end{itemize}

Thus, the $(n,m)$ Pad\'e approximant of a given analytic function is defined as
	\begin{equation}
	P_{n,m}(z)=\dfrac{\displaystyle \sum_{i=0}^n a_i z^i}{1+\displaystyle\sum_{k=1}^m b_k z^k}\,,
	\end{equation}
where $n$ and $m$ are the degrees of the polynomials at the numerator and denominator, respectively, and $a_i$ and $b_k$ are constant coefficients to be determined after fixing the  expansion to the desired order \cite{Baker96}.

Hence, we construct the Pad\'e parametrization of $q_{de}(z)$ such that
	\begin{equation}
	q_{de}=-1+\dfrac{(1+z)}{H_{de}}\dfrac{dH_{de}}{dz}\,,
	\label{eq:q_de}
	\end{equation}
where $H_{de}^2=\frac{1}{3}\rho_{de}$ is the Hubble expansion rate of the dark energy term. 

A suitable Pad\'e approximant to be used in order to construct a total $q(z)$ able to fulfill the conditions (\ref{constraint1}), (\ref{constraint2}) and (\ref{constraint3})  is the (0,1) parametrization, so that
    \begin{equation}
    q_{de}(z)=\dfrac{q_{de,0}}{1+q_1 z}\,,
    \label{eq:q_de2}
    \end{equation}
where $q_{de,0}$ is the corresponding value at the present time $(z=0)$.
Our choice is motivated since:

\begin{itemize}
    \item[-] we require dark energy to be subdominant over matter at very large redshifts and, indeed, we here have $q_{de}\rightarrow0$ as $z\rightarrow\infty$;
    \item[-] we need the smallest order expansion to check the goodness of our parametrization as the simplest demands are considered. 
\end{itemize}

It is worth noticing  that we started focusing on the dark energy deceleration parameter rather than the total one computed from the full Hubble rate, comprising matter, dark energy etc. This permits us to impose the basic hypotheses on suitable dark energy evolution only, while the conditions stressed in  \eqref{constraint1}-\eqref{constraint3} will be guaranteed on the total deceleration parameter. 

Therefore, using \Cref{eq:q_de2} in \Cref{eq:q_de}, one obtains 
	\begin{equation}
	H_{de}=\mathcal C(1+z)^{1+\frac{q_{de,0}}{1-q_1}}(1+q_1z)^{-\frac{q_{de,0}}{1-q_1}}\,,
	\end{equation}
where $q_1\neq 1$, and the constant of integration $\mathcal C$ is such that the present value of $H= H_{m}+H_{de}$ coincides with the Hubble constant $H_0$, namely 

\begin{equation}
    \mathcal C=H_0\sqrt{1-\Omega_{m0}}\,.
\end{equation}

Consequently, the dark energy density is given by $\rho_{de}=3H_{de}^2$, and including the contribution of matter, $\rho_m=3H_0^2\Omega_{m0}(1+z)^3$, we have
	\begin{align}
	&H^2=H_0^2\bigg[\Omega_{m0}(1+z)^3+(1-\Omega_{m0})(1+z)^{2+\frac{2q_{de,0}}{1-q_1}} \nonumber  \\
	&\hspace{4.5cm}\times(1+q_1z)^{-\frac{2q_{de,0}}{1-q_1}}\bigg]. \label{eq:H2}
	\end{align}
The total deceleration parameter is thus given as
	\begin{equation}
	q=\dfrac{2q_{de,0}(1-\Omega_{m0})(1+q_1z)^{-1+\frac{2q_{de,0}}{q_1-1}}+\Omega_{m0}(1+z)^{1+\frac{2q_{de,0}}{q_1-1}}}{2\left[(1-\Omega_{m0})(1+q_1z)^{\frac{2q_{de,0}}{q_1-1}}+\Omega_{m0}						(1+z)^{1+\frac{2q_{de,0}}{q_1-1}}\right]}
	\end{equation}
Requiring the above expression to obey the condition  (\ref{constraint3}), we find the constraint $q_1=1+q_{de,0}$, leading to
	\begin{equation}
	q(z)=\dfrac{2q_{de,0}(1-\Omega_{m0})(1+z+q_{de,0}z)+\Omega_{m0}(1+z)^3}{2\left[(1-\Omega_{m0})(1+z+q_{de,0}z)^2+\Omega_{m0}(1+z)^3\right]}\,.
	\end{equation}
Remarkably, the obtained expression satisfies also the constraint (\ref{constraint1}), while we shall verify the condition (\ref{constraint2}) through the constraints from cosmic observables.

Finally, imposing $q_1=1+q_{de,0}$ in \Cref{eq:H2}, we find
    \begin{equation}
    H(z)=H_0\sqrt{\Omega_{m0}(1+z)^3+(1-\Omega_{m0})(1+z+q_{de,0}z)^2}\,.
    \label{eq:H}
    \end{equation}
It is interesting to note that our result represents a one-parameter extension of the $\Lambda\text{CDM}$ model, which is exactly recovered in the limit $q_{de,0}\rightarrow-1$. 

\section{Linear perturbations}

Let us also investigate the behaviour of the above model in the regime of linear perturbations. These are described by density fluctuations of matter on sub-horizon scales in the Newtonian approximation \cite{structures}:
\begin{equation}
\ddot{\delta}_m+2H\dot{\delta}_m-\dfrac{1}{2}\rho_m\delta_m=0\,,
\end{equation}
where $\delta_m\equiv \delta\rho_m/\rho_m$ is the matter density contrast, and the dot denotes derivative with respect to the cosmic time. After some manipulations, the previous equation can be written in terms of the scale factor as 
	\begin{equation}
	\delta_m''+\left(\dfrac{3}{a}+\dfrac{E'}{E}\right)\delta_m'-\dfrac{3\Omega_{m0}}{2a^5E^2}\delta_m=0\,,
	\end{equation}
where, in our case, $E(a)$ is obtained from \Cref{eq:H} being $E(a(z))\equiv H(z)/H_0$ and substituting  $a=(1+z)^{-1}$.

The analysis of the growth of matter perturbations is usually addressed by means of the quantity $f(a)\equiv a\delta'_m(a)$ and the factor $f\sigma_8(a)\equiv f(a)\sigma_8(a)$, where 
$\sigma_8(a)=\sigma_8	\delta_m(a)/\delta_m(1)$ represents the \emph{rms} fluctuations of the linear density field within a $8h^{-1}$ Mpc radius, with $\sigma_8$ being its present value \cite{Planck18,Lambiase}.

\section{Observational analysis}
\label{sec:observations}

The proposed parametrization can be tested against cosmic data. We  shall use the observational constraints to explore the consequences on dark energy.

\subsection{Markov Chain Monte Carlo results}

\begin{table*}
\begin{center}
\setlength{\tabcolsep}{1em}
\renewcommand{\arraystretch}{2}
\begin{tabular}{c c c c c c c c c}
\hline
\hline
Model & $h_0$ & $\Omega_{m0}$ & $q_{de,0}$  & $\sigma_8$ \\
\hline
New DE & $0.697\pm 0.019\,(0.038)   $ & $0.283^{+0.054\,(0.093)}_{-0.049\,(0.104)}$  & $-1.02^{+0.21\,(0.38)}_{-0.21\,(0.42)} $  &  $0.791^{+0.066\,(0.189)}_{-0.096\,(0.159)}   $ \\
$\Lambda$CDM & $0.698\pm 0.018\,(0.036)$ &  $0.282^{+0.024\,(0.048)}_{-0.024\,(0.045)}   $ & $-1$ & $0.782^{+0.035\,(0.071)}_{-0.035\,(0.067)}$  \\
\hline
\hline
\end{tabular}
\caption{$68\%\,(95\%)$ confidence level results from the MCMC analysis of the new dark energy parametrization. The last two columns show the results from the information criteria, calculated with respect to the reference $\Lambda\text{CDM}$ scenario. It is worth to note that $q_{de,0}$ here refers to the present value of the dark energy deceleration parameter, instead of the total one that fulfills the requirement of being larger than $-1$ \cite{cosmografia2012}.}
 \label{tab:results}
\end{center}
\end{table*}

To analyze the background evolution, we take into account the low-redshift model-independent observational Hubble data from the differential age method \cite{Jimenez02} and the Supernovae Ia data of the \emph{Pantheon} catalogue \cite{Pantheon}, in the redshift interval $0<z<2$. 
On the other hand, matter perturbations can be addressed through the use of the  \emph{Gold-2017}  compilation \cite{Nesseris17} of measurements from weak lensing and redshift space distortion.

We thus combine  data to build up the joint likelihood, and we use the Markov Chain Monte Carlo (MCMC) method to sample the parameter space defined by the proposed scenario to obtain numerical constraints over the free parameters. We refer to \cite{D'Agostino19} for the details on the measurements and their relative uncertainties, as well as for the likelihood functions of the single datasets.

\begin{figure}
\includegraphics[width=3.3in]{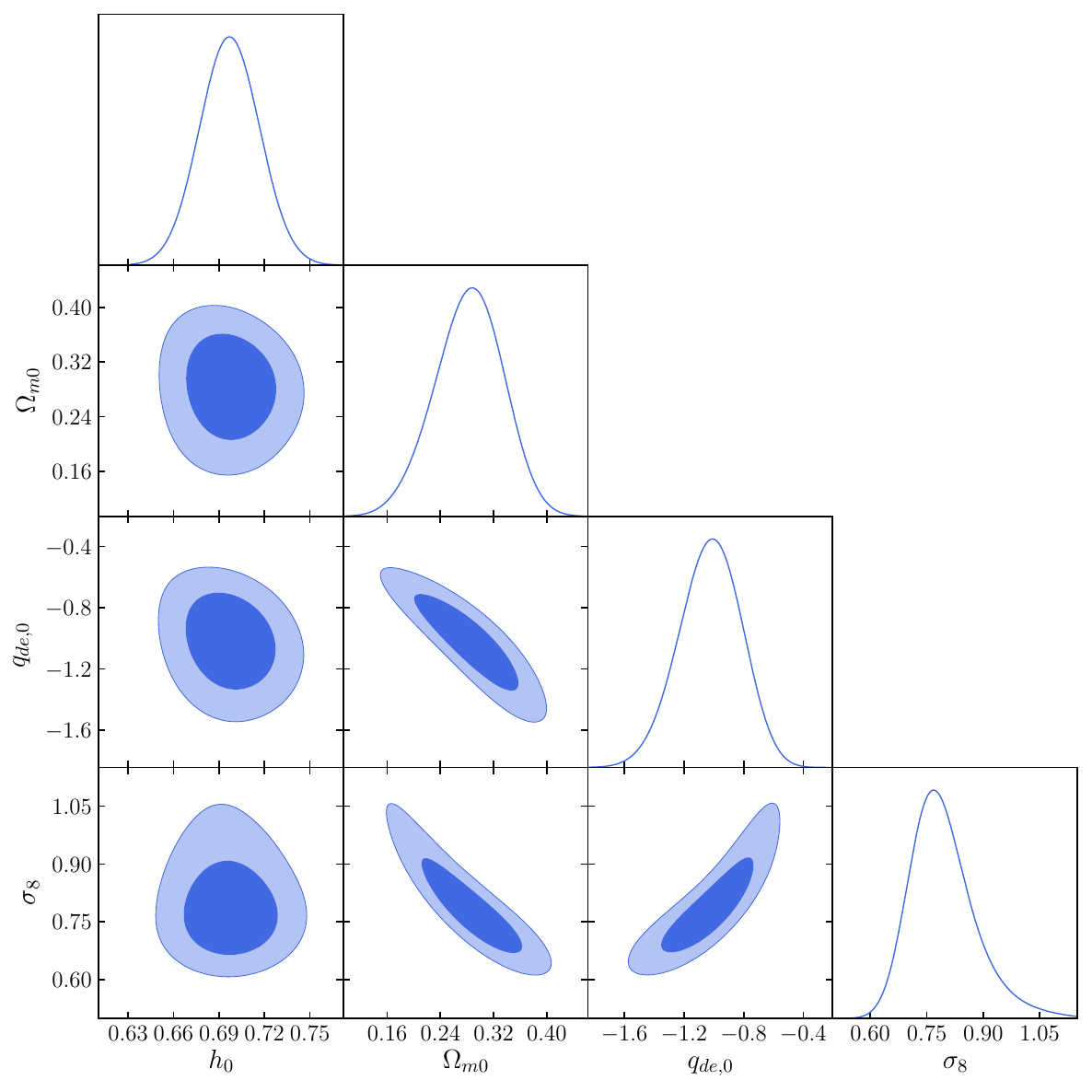}
\caption{MCMC contours at the 68\% and 95\% confidence levels and posterior distributions of the free coefficients of the new dark energy parametrization.}
\label{fig:contours}
\end{figure}

The results up to the $2\sigma$ confidence level are summarized in \Cref{tab:results}, where we also show. for comparison, the predictions of the standard $\Lambda$CDM model.
Moreover, \Cref{fig:contours} shows the marginalized $1\sigma$ and $2\sigma$ contour regions with posterior distributions for the free parameters.  Results on the Hubble constant are given in terms of the dimensionless parameter $h_0\equiv H_0/(100\ \text{km/s/Mpc})$. 

Our analysis permits to test the deviations from the $\Lambda$CDM model. In particular,  constraints on the $q_{de,0}$ parameter show that no significant evidence against the standard model is present at the $1\sigma$ level. Also, the matter density is fully consistent within $1\sigma$ with the most recent value inferred by the \emph{Planck} collaboration \cite{Planck18} from the anisotropies of the cosmic microwave background.

Furthermore, the value of the Hubble constant, resulting from our analysis, differs by $1.8\sigma$ from the direct measurement by Riess et al. \cite{Riess19}, while being $1.2\sigma$ away from the \textit{Planck} findings \cite{Planck18}. 
Our cosmographic approach thus alleviates the tension between the local and CMB estimates of $H_0$, showing close similarities to the predictions of other parametric models such as emergent dark energy proposed in \cite{EDE}, where the dark energy effects emerge only at late times while not being effectively present at earlier stages of the cosmic evolution.

\subsection{Consequences on dark energy}

\begin{figure}
\begin{center}
\includegraphics[width=3.2 in]{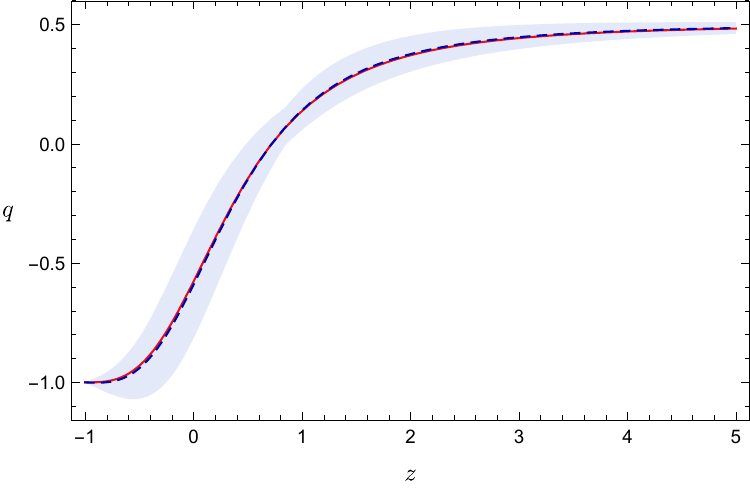}
\caption{Reconstruction of the deceleration parameter as a function of the redshift based on the MCMC analysis of the new dark energy parametrization. The shaded regions around the mean result (dotted blue) account for the $1\sigma$ uncertainties. The $\Lambda\text{CDM}$ mean result (solid red) is shown for comparison.}
\label{fig:q}
\end{center}
\end{figure}

The results of our MCMC analysis can be used to study the dark energy behaviour. 
In particular, in \Cref{fig:q},  we show the reconstruction of the deceleration parameter, compared to the prediction of  the  $\Lambda$CDM model. We note that the transition between the decelerating and accelerating phases of the Universe evolution $(q=0)$ occurs at $z_t\simeq0.72$, in agreement with the model-independent findings previously obtained in \cite{CDLandVargas16}.

\begin{figure*}
\centering
\subfigure{\includegraphics[width=3.2in]{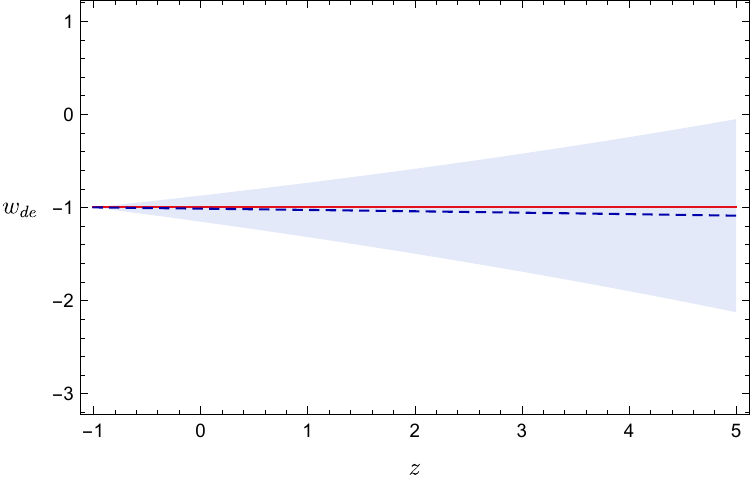}}
\hspace{1cm}
\subfigure{\includegraphics[width=3.2in]{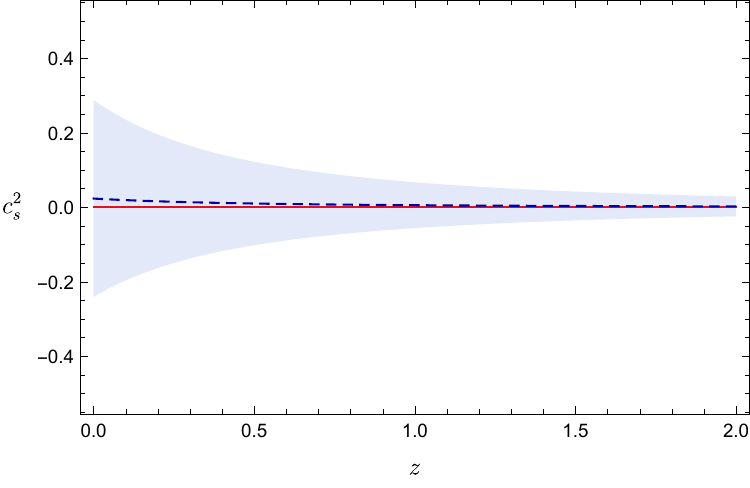}}
\caption{Dark energy EoS parameter (left panel) and sound speed (right panel) as functions of the redshift inferred from the MCMC analysis of the new dark energy parametrization compared to cosmological constant case (solid red). The shaded regions around the mean result (dotted blue) account for the $1\sigma$ uncertainties.}
\label{fig:w-cs}
\end{figure*}

Furthermore, to investigate the properties of cosmic acceleration, we consider a generic dynamical dark energy model, for which
	\begin{equation}
	E(z)^2=\Omega_{m0}(1+z)^3+(1-\Omega_{m0})X(z)\,,
	\end{equation}
where
	\begin{equation}
	X(z)=\exp{\left\{3\int_0^z \dfrac{1+w_{de}(z')}{1+z'}\, dz' \right\}}\,.
	\end{equation}	
Here, $w_{de}(z)$ is the dark energy equation of state (EoS) parameter. It is then easy to show that
	\begin{equation}
	w_{de}(z)=\dfrac{\frac{2}{3}(1+z)\frac{d\ln E(z)}{dz}-1}{1-\frac{\Omega_{m0}(1+z)^3}{E(z)^2}}\,.
	\label{eq:w}
	\end{equation}
In our case, using the Hubble expansion rate given by \Cref{eq:H}, we obtain
	\begin{equation}
	w_{de}(z)=\dfrac{1}{3}\left[\dfrac{2q_{de,0}}{1+z(1+q_{de,0})}-1\right] . 
	\end{equation}
We note that, for $q_{de,0}=-1$, we have $w_{de}=-1$ corresponding to the cosmological constant.
The left panel of \Cref{fig:w-cs} shows the behaviour of the dark energy EoS parameter in view of  the results of the MCMC analysis. We see that the mean curve suggests only a very weak dynamical feature for dark energy, which is however well consistent with the $\Lambda$CDM scenario at the $1\sigma$  level.

A physical quantity playing a key role in cosmic structure formation is the sound speed \cite{Mukhanov92}, given by
	\begin{equation}
	c_s^2\doteq\dfrac{d p}{d \rho}=-1-\dfrac{\ddot{H}}{3H\dot{H}}\,,
	\end{equation}
where the last equality holds by virtue of the Friedman equations. Thus, using \Cref{eq:H}, we find
	\begin{equation}
	c_s^2=\dfrac{1+z(1+q_{de,0})-q_{de,0}}{-3\left[1+z(1+q_{de,0})+\frac{3\Omega_{m0}(1+z)^2}{2(1-\Omega_{m0})(1+q_{de,0})}\right]}	\,.
	\end{equation}
Once again, it is easy to verify that, for $q_{de,0}=-1$, we recover the $\Lambda$CDM prediction $(c_s^2=0)$. The behaviour of the sound speed, as a result of the MCMC analysis, is shown in the right panel of \Cref{fig:w-cs}.

\begin{figure}
\includegraphics[width=3.2in]{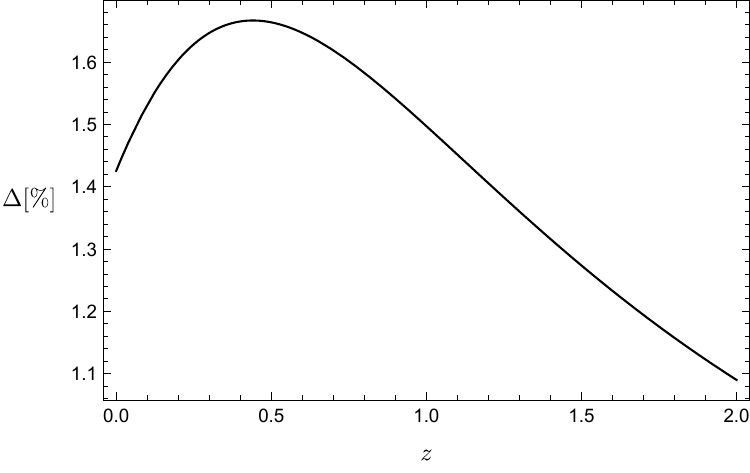}
\caption{Relative percentage difference of the growth rate factor of the new dark energy parametrization with respect to the $\Lambda$CDM model, corresponding to the mean results from the MCMC analysis.}
\label{fig:Delta}
\end{figure}

Finally, to explore the consequences at the perturbation level, we analyze the growth rate factor and its relative difference with respect to the standard cosmological model by means of  the quantity
\begin{equation}
\Delta(z)	\equiv \dfrac{f\sigma_8(z)-f\sigma_{8,\Lambda\text{CDM}}(z)}{f\sigma_{8,\Lambda\text{CDM}}(z)}\,.
\end{equation}
Using the mean results from the MCMC analysis, we can see, in \Cref{fig:Delta}, that the deviations from $\Lambda$CDM are $<2\%$ in the redshift interval of the observational data.

\section{Outlook and perspectives}
\label{sec:conclusion}

In this paper, we built up a dark energy parametrization that works fairly well in mimicking the dark sector at both late and early times. Making use of thermodynamic requirements and kinematic constraints over the Universe evolution, we approximated the dark energy deceleration parameter  by means of Pad\'e rational polynomials. We motivated our choice considering a dark energy term subdominant over matter and radiation as $z\rightarrow\infty$. In fulfillment with thermodynamic demands, we assumed the second law of thermodynamics to hold in a spatially-flat FLRW Universe, guaranteeing $dS>0$ and $d^2S<0$ as $z\rightarrow-1$,  having $q\rightarrow -1 \ \left(\dfrac{dq}{dz}>0\right)$ as $z\rightarrow -1$,  and $q\geq-1$ $  \forall z$, with $q\rightarrow \dfrac{1}{2}$ for $z\gg 1$. We showed that, neglecting radiation, the corresponding total deceleration parameter, accounting for matter and dark energy fluids, works well at both late and early times. 

For this purpose, we performed a numerical analysis through a Monte Carlo procedure based on Metropolis algorithm, to obtain constraints on the free parameters of the model and to confront the behaviour of the parametrization with the standard cosmological scenario. The same has been worked out for linear perturbations, showing that the sound speed is always positive definite, letting structures to form accordingly to observations. Moreover, we compared the growth rate of matter fluctuations resulting from our model with the predictions of the $\Lambda$CDM model.

The results of the study show only small differences between the standard cosmological model and our approach. Thus, we can conclude that any dark energy parametrization, made up through suitable barotropic fluids and/or extensions of General Relativity, might fulfill the requirements we showed, within the small departures prompted from our findings.

Future works will explore alternative dark energy parametrizations with additional thermodynamic requirements. Moreover, the approach can be improved using  different rational approximations of the deceleration parameter to check whether this would be helpful in disclosing early time cosmology. Furthermore, a particular attention can be devoted to investigate the Hubble constant tension in view of tighter observational constraints.

\acknowledgments 
S.C. and R.D. acknowledge the support of INFN (\emph{iniziativa specifica} QGSKY).
O.L. acknowledges the support of the Ministry of Education and Science of the Republic of Kazakhstan, Grant IRN AP08052311.

\end{document}